# 3D Printable Plasmonic Titanium Nitride Nanoparticles Enhanced Thermoplastic Polyurethane Composite for Improved Photothermal De-Icing and Infrared Labeling

Siyu Lu[a], Jixiang Zhang[a], Min Xi*[b], Nian Li*[b], Zhenyang Wang*[b]

[a] School of Mechatronics and Vehicle Engineering, Chongqing Jiaotong University, Chongqing, 400074, P. R. China

[b] Institute of Solid State Physics and Key Laboratory of Photovoltaic and Energy Conservation Materials, Hefei Institutes of Physical Science, Chinese Academy of Sciences, Hefei, Anhui 230031, P. R. China

Corresponding authors:

Min Xi − orcid.org/0000-0003-4414-3110; Email: minxi@issp.ac.cn

Nian Li; Email: linian@issp.ac.cn

Zhenyang Wang − orcid.org/0000-0002-0194-3786; Email: zywang@iim.ac.cn

## Table of Contents

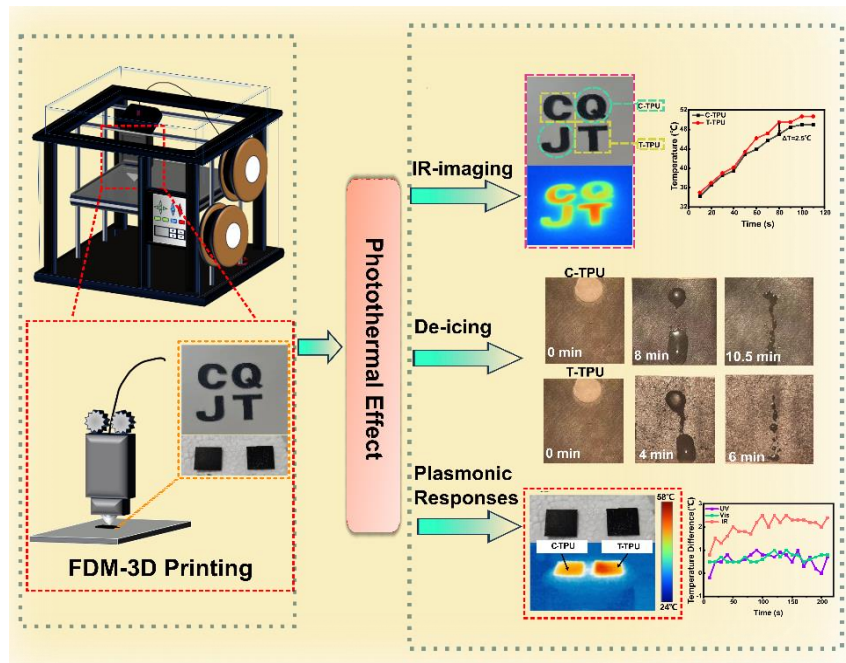

A titanium nitride (TiN) enhanced composite filament (T-TPU) was developed for 3D printing, that showed plasmonic response to different bands of light and exceptional photothermal performance in de-icing and infrared labeling

## Abstract

Plasmonic nanomaterials offer a direct and effective approach to harnessing solar energy. Specifically, plasmonic semiconductors enable a highly efficient light-to-heat conversion process, outperforming noble metals in stability, cost-effectiveness, and accessibility. In this study, a composite 3D printing filament (T-TPU), composed of titanium nitride (TiN) and thermoplastic polyurethane (TPU), was synthesized using a combined extrusion process involving a twin-screw extruder and a single-screw extruder. The resulting T-TPU filament could be used with fused deposition modeling (FDM) 3D printing to produce custom-designed patterns. Notably, these printed patterns exhibited superior photothermal performance, with potential applications in photothermal de-icing and infrared labeling. Additionally, the wavelength-dependent plasmonic and photothermal responses of the printed patterns were experimentally investigated and supported by finite elemental method (FEM) simulations, revealing a temperature increase of approximately 2.5 °C under IR LED light when compared to commercial black thermoplastic polyurethane (C-TPU), that was more obvious than a difference less than 1 °C under UV or visible LED light sources. Finally, the mechanical properties of T-TPU, altered by the inclusion of TiN nanoparticles, were assessed, showing a slight enhancement in modulus and friction coefficient relative to neat TPU (N-TPU). Molecular dynamics (MD) simulations indicated that the TiN nanoparticles promoted strong interactions between polymer chains and TiN particles, enhancing the modulus of elasticity and contributing to the improved mechanical properties of T-TPU. These findings suggest improved abrasion resistance, demonstrating the stability and durability of the composite material.

## Introduction

Plasmonic nanomaterials have demonstrated highly efficient photothermal conversion,[1,2] enabling applications such as solar steam generation (SSG),[3] photonic pathogen inactivation,[4,5] cell resuscitation and rewarming,[6,7] etc. Notably, plasmonic nanomaterials serve as effective solar absorbers, providing a promising approach to anti-icing and de-icing.[8–10] The development of photothermal coatings holds significant potential for aircraft applications, where preventing ice formation is crucial for safe operation in extreme weather.[11] Additionally, plasmonic photothermal nanomaterials offer unique possibilities for infrared labeling, valuable in anti-counterfeiting as physically unclonable functions (PUFs) with distinct optical properties.[12] Compared to traditional anti-counterfeiting methods, PUF marks offer advantages such as material specificity, randomness, and variability, making them resistant to replication or imitation.[13–15]

The efficacy of these photothermal applications hinges on the heat collection efficiency of the underlying structures. Achieving efficient heat collection requires a photothermal medium that is productive, stable, and cost-effective.[16,17] As a semiconducting plasmonic material, titanium nitride (TiN) nanoparticles have gained attention as an alternative to gold nanoparticles in various fields due to their adjustable lattice parameters and their optical and plasmonic response across the visible and near-infrared spectrum.[18–20] In particular, TiN nanoparticles outperform gold in photothermal applications due to their relatively lower electron concentration, reduced energy loss, abundant availability, and high chemical and physical stability, positioning TiN as an ideal candidate for photothermal technologies.[21–24]

Efforts to incorporate TiN nanoparticles into nanocomposites for diverse applications have employed methods such as particle deposition,[25] polymer coating,[26] self-assembly,[27] inkjet printing,[28] and 3D printing.[29,30] Specifically, fused deposition modeling (FDM),[31] an extrusion-based 3D printing technique, offers a direct and versatile approach to personalized design and rapid production of photothermal patterns, adding an extra dimension of customization in the fabrication of these functional structures.[32,33]

In this study, a TiN nanoparticle-reinforced thermoplastic polyurethane (T-TPU) composite filament was synthesized and subsequently used for 3D printing of custom-designed photothermal patterns via fused deposition modeling (FDM). Compared to neat thermoplastic polyurethane (N-TPU), T-TPU exhibited enhanced thermal properties, such as enhanced thermal conductivity (0.52 W/(m·K) of T-TPU with 5 wt% TiN compared to 0.26 W/(m·K) of N-TPU), increased heat capacity (1.99 J/(g·°C) of T-TPU with 5 wt% TiN compared to 1.75 J/(g·°C) of N-TPU) and slightly increased density (1.76 g/cm$^3$ of T-TPU with 5 wt% TiN to compared to 1.21 g/cm$^3$ of N-TPU. These changes of thermal properties would benefit the photothermal performance of fabricated patterns. Solar photothermal de-icing and IR labeling experiments were conducted to validate its photothermal performance. Under 0.15 W/cm² simulated sunlight, T-TPU patterns reached equilibrium temperatures approximately 2.5 ℃ higher than those of commercial black colored TPU (C-TPU) within 2 minutes. The T-TPU's rapid heat collection significantly improved de-icing performance, with 3 g of ice melting in 6 minutes on T-TPU under 0.15 W/cm² sunlight, compared to 10.5 minutes on C-TPU. Additionally, the plasmonic response of T-TPU under different wavelengths was investigated. T-TPU patterns displayed a ~2.0 ℃ higher temperature than C-TPU under IR (805 nm, 0.10 W/cm²) illumination, while no significant temperature difference was observed under UV (360 nm, 0.10 W/cm²) or visible (525 nm, 0.10 W/cm²) radiation that was comparable to the reported ITO nanoink performance.[34] Numerical simulations confirmed a structure–effect relationship between filament optical density and the temperature elevation of printed patterns. Finally, the mechanical properties of T-TPU were examined. The addition of TiN nanoparticles resulted in a slight decrease in the friction coefficient and modulus. Molecular dynamics simulations indicated that the TiN nanoparticles promoted strong interactions between polymer chains and TiN particles, enhancing the modulus of elasticity and contributing to the improved mechanical properties of T-TPU.

## Results and Discussion

### 1. Preparation of T-TPU Composite Filaments

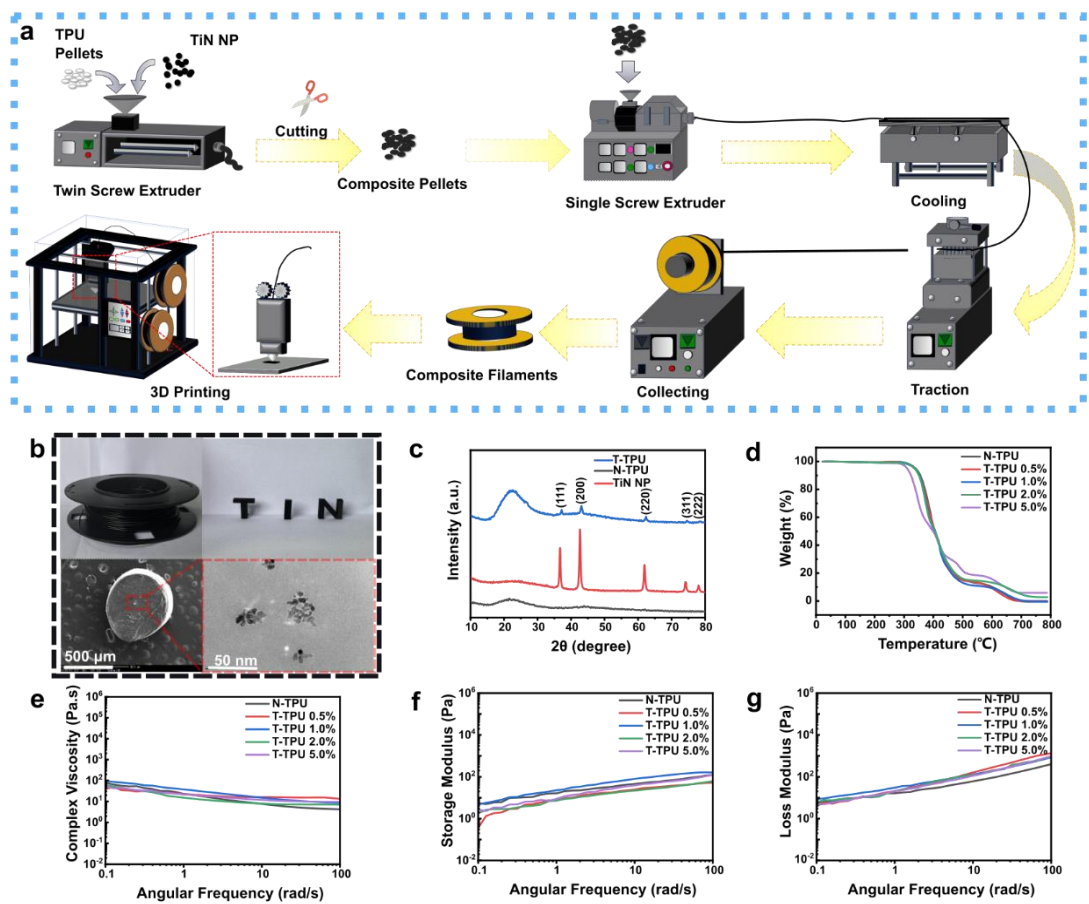

**Figure 1** Synthesis and characterization of T-TPU 3D printing filament. **(a)** Schematic of the T-TPU 3D printing filament fabrication process; **(b)** Photographs of the synthesized T-TPU filament (upper left) and printed models using the T-TPU filament (upper right), with SEM image showing the cross-section of the extruded filament (lower left) and TEM image illustrating the TiN nanoparticle distribution within the filament (lower right); **(c)** XRD patterns for 5 wt% T-TPU, neat TPU (N-TPU), and TiN nanoparticle powders; **(d)** Thermogravimetric analysis (TGA) curves of T-TPU composite filaments with varying TiN content; **(e) – (g)** Complex viscosity, storage modulus, and loss modulus of T-TPU composite filaments as functions of scanning frequency, showing the influence of different TiN concentrations.

As shown in **Figure 1a**, the TiN nanoparticle-reinforced thermoplastic polyurethane (T-TPU) composite filament was synthesized via a modified two-step process based on previously reported methods. In this approach, TiN nanoparticles (**Figure S1**, characterized by HR-TEM) were melt-blended with TPU granules in a twin-screw extruder to achieve a homogeneous distribution. The blended composite was then extruded, cooled, and manually cut into T-TPU masterbatch granules. These granules were subsequently processed into a wire-shaped filament using a single-screw extruder, with extrusion parameters such as screw speed and temperature carefully controlled. The final composite filament, shown coiled in the upper left inset of **Figure**

**1b**, had a diameter of 1.75 ± 0.1 mm, allowing for efficient collection by a winding machine.

This processed filament was then used for fused deposition modeling (FDM) printing, where its printability was validated by the high accuracy of printed letters (upper right inset, **Figure 1b**), demonstrating its capability to produce complex, precise structures. The lower left inset of **Figure 1b**, an SEM image of the filament cross-section extruded from the 3D printer nozzle, revealed a smooth surface free from irregularities and distortions, even with 5 wt% TiN nanoparticles, ensuring smooth and uninterrupted printing. The lower right inset of **Figure 1b**, a TEM image of a filament cross-section, confirms the successful incorporation of TiN nanoparticles within the TPU matrix. Additionally, the uniform distribution of TiN nanoparticles in the T-TPU filament was further supported by the X-ray diffraction (XRD) pattern in **Figure 1c**, displaying characteristic TiN peaks at (111), (200), (220), (311), and (222).

As demonstrated by the thermogravimetric analysis (TGA) results in **Figure 1d**, the TiN content in the T-TPU filament could be controlled by adjusting the feed ratio, with residual weights closely matching nominal weight compositions, indicating minimal filler loss during filament preparation.[35] **Figure 1e - 1g** illustrated the complex viscosity, storage modulus (G'), and loss modulus (G") of T-TPU filament samples with varying TiN content across different scanning frequencies. **Figure 1e** showed a decrease in complex viscosity with increasing frequency, characteristic of shear thinning in polymer melts, which supported smooth extrusion.[36] The comparatively low viscosity of the molten T-TPU further enabled homogeneous distribution of the TiN nanoparticles.[37]

**Figures 1f** and **1g** illustrated that both the storage modulus (G') and loss modulus (G") increased steadily, forming a frequency-dependent plateau, indicative of excellent printability for extrusion-based 3D printing of the T-TPU composite filament. The TiN content significantly influenced the thermal properties of the filament, as variations in TiN dispersion led to changes in these properties. For instance, a printed plate model (20 × 20 × 2 mm³, **Figure S2**) was used to assess thermal characteristics, showing that with increasing TiN weight percentage, the thermal diffusivity, density, heat capacity, and thermal conductivity of the T-TPU filaments also increased (**Figures S3a – S3d**), such as enhanced thermal conductivity (0.52 W/(m·K) of T-TPU with 5 wt% TiN compared to 0.26 W/(m·K) of N-TPU), increased heat capacity (1.99 J/(g·°C) of T-TPU with 5 wt% TiN compared to 1.75 J/(g·°C) of N-TPU) and slightly increased density (1.76 g/cm$^3$ of T-TPU with 5 wt% TiN to compared to 1.21 g/cm$^3$ of N-TPU. This enhancement was likely due to the increased filler content, which reduced voids and defects in the TPU matrix, creating additional heat transfer pathways.

## 2. Research on Thermal and Photothermal Performance

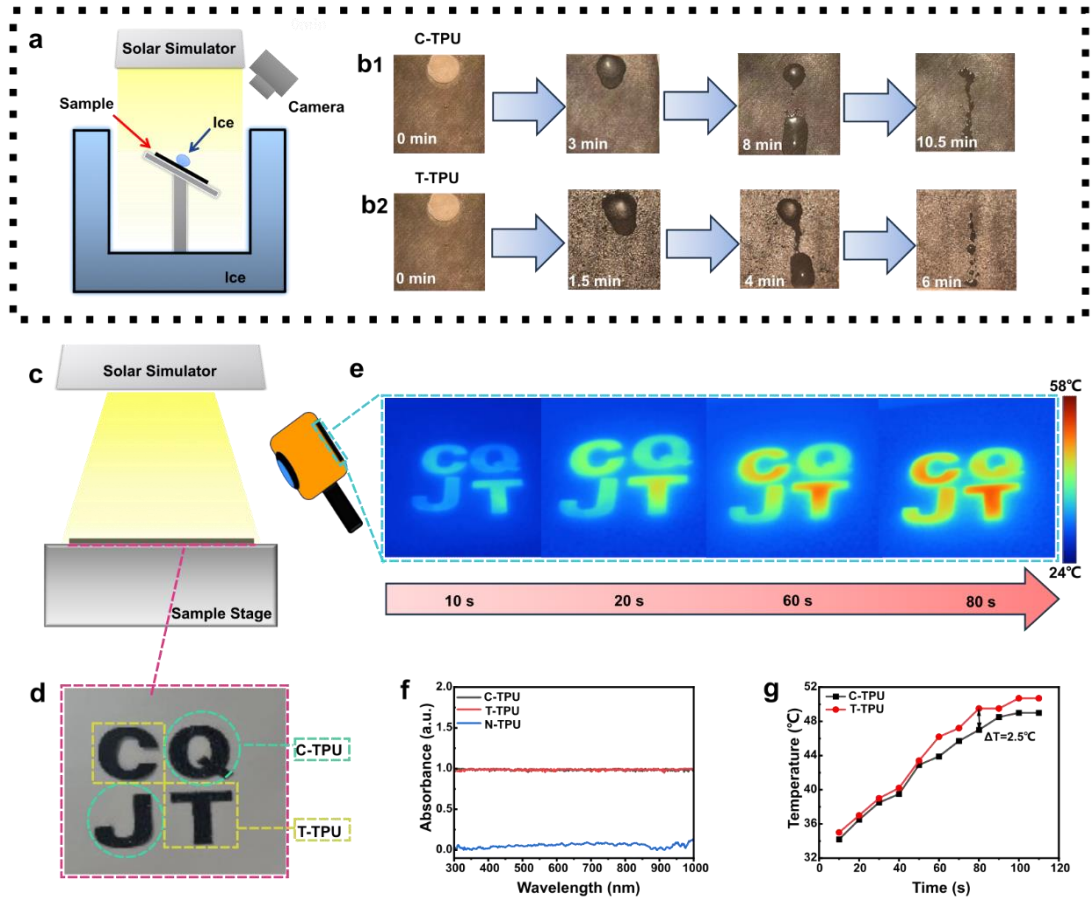

**Figure 2** Photothermal de-icing performance and photothermal IR imaging study. **(a)** Schematic of the experimental setup for the photothermal de-icing test; **(b1)** and **(b2)** Representative images illustrating the photothermal de-icing performance of 3D-printed C-TPU and T-TPU samples over time, respectively; **(c)** Schematic of the experimental setup for the photothermal IR imaging test; **(d)** Images of 3D-printed samples used in the photothermal IR imaging test; **(e)** Representative infrared images of printed letter samples under simulated solar illumination; **(f)** UV-Vis-NIR spectra of films (1 mm thickness) printed with C-TPU, N-TPU, and T-TPU composite filaments; **(g)** Temperature variation of letters printed with C-TPU and T-TPU filaments under simulated solar illumination as a function of time.

As discussed, the synthesized T-TPU filament exhibited excellent printability, and the incorporation of TiN nanoparticles enhanced its thermal properties, making it well-suited for photothermal applications. The photothermal de-icing performance was first evaluated. Using a 3D modeling software and an FDM fabrication process, two layers that one printed with commercially available black colored TPU (C-TPU) and another with T-TPU were prepared at a thickness of 1 mm each. As shown in **Figure 2a**, 3 mL of water was placed on the sample surfaces and frozen at -80 °C. The samples were then positioned on a 30° inclined stage in a 0 °C environment with ice surrounding the area. A simulated solar light source with an intensity of 0.15 W/cm² illuminated the sample surfaces, and the melting process was recorded. **Figure 2b1** showed that ice on the C-TPU sample began melting after 3 minutes, with half of the ice volume gone by 8 minutes. Complete melting occurred by 10.5 minutes, leaving minor water traces. In contrast, as shown in **Figure 2b2**, ice on the T-TPU sample began melting within 1.5 minutes, with over half melted by 4 minutes and complete melting by 6 minutes, that was a reduction in de-icing time of over 40%. This significant improvement was attributed to the TiN nanoparticles' photothermal conversion efficiency and broad-band solar absorption, coupled with their uniform distribution in the TPU matrix, enabling rapid and even heat dissipation within the T-TPU layer.

To further evaluate the T-TPU composite filament's potential for photothermal infrared imaging, an infrared imaging test was conducted, as depicted in **Figure 2c** and **2d**. Alphabetical models were created (**Figure S4**) and printed via FDM, with letters "C" and "T" printed in T-TPU and "Q" and "J" in C-TPU filament (**Figure 2d**). The samples were then illuminated by a solar simulator (0.15 W/cm²), while a portable infrared thermal camera captured temperature changes. **Figure 2e** showed representative IR images, illustrating a shift from cool to warm tones over 80 seconds of illumination, although both types of letters appeared visually indistinguishable to the naked eye. To evaluate optical transparency, 20 × 20 × 1 mm³ films were printed with each filament (C-TPU, T-TPU, and N-TPU) and analyzed for absorbance in the range from 300 nm to 1000 nm (**Figure 2f**). Both C-TPU and T-TPU films showed nearly 100 % absorbance across this range, while N-TPU remained nearly transparent in the visible and NIR range. Remarkably, T-TPU samples consistently exhibited a more intense IR image compared to C-TPU samples (**Figure 2e**). Temperature analysis of the IR images, conducted with SmartView Classic 4.4 software, confirmed that T-TPU samples maintained higher temperatures than C-TPU samples, with a peak difference of ~2.5 °C at 80 seconds (**Figure 2g**). The nano-sized TiN nanoparticles contributed significantly to improved thermal conductivity and photothermal conversion, resulting in the T-TPU's superior performance. The distinct infrared contrast and temperature difference highlight the T-TPU composite's potential for heat collection, infrared imaging, and anti-counterfeiting applications.

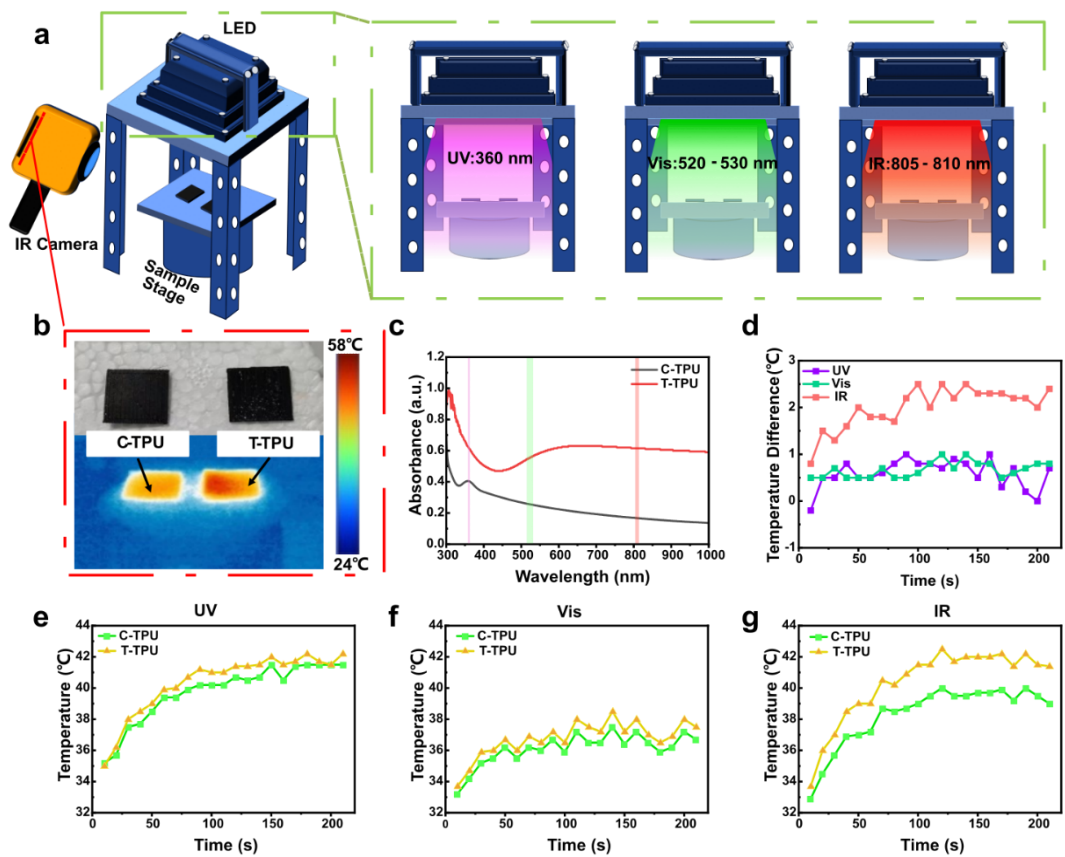

**Figure 3** Photothermal performance under different wavelength light sources. **(a)** Schematic of the experimental setup; **(b)** 3D-printed samples for photothermal-IR imaging with corresponding infrared images; **(c)** UV-Vis-NIR spectra of T-TPU and C-TPU composites at a concentration of 2 mg/mL in DMF, with labeled LED light source wavelengths; **(d)** Temperature differences between T-TPU and C-TPU layers over time under UV, green, and IR light sources; **(e) - (g)** Temperature variations of T-TPU and C-TPU layers over time for each light source of UV, visible, and IR, respectively.

Due to the plasmonic nature of titanium nitride (TiN) nanoparticles, it was anticipated that the T-TPU composite would exhibit varying photothermal responses across different wavelength bands. As illustrated in **Figure 3a**, three LED lamps with centered wavelengths of 360 nm (UV), 520-530 nm (visible), and 805-810 nm (near-infrared) were selected to investigate the photothermal behavior of the T-TPU composite. The LED light sources, each emitting an illumination power of 0.1 W/cm², were positioned above the 3D-printed T-TPU and C-TPU layers, while a portable infrared camera was utilized to capture infrared images and monitor temperature fluctuations.

As depicted in **Figure 3b**, the printed layer samples of C-TPU and T-TPU, each measuring 20 × 20 × 1 mm, appeared nearly identical in color but emitted distinct infrared (IR) light, indicating differing photothermal conversion efficiencies. This photothermal response could be intuitively linked to the optical density of the composites. The absorbance spectra of C-TPU and T-TPU composite filaments dissolved in dimethylformamide (DMF) at a concentration of 2 mg/mL were shown in **Figure 3c**. The T-TPU composite exhibited a higher optical density than the C-TPU composite, particularly demonstrating a broad plasmonic peak around 700 nm, in addition to high absorption around 300 nm in the deep UV, that was consistent with previous literature. The absorbance values for T-TPU at 360 nm, 525 nm, and 805 nm were 0.62, 0.56, and 0.61, respectively, compared to C-TPU's 0.40, 0.26, and 0.17, resulting in ratios of 1.55, 2.15, and 3.59.

**Figure 3d - 3f** summarized the temperature variations of the C-TPU (green dots and curves) and T-TPU (yellow dots and curves) layers based on the recorded IR images. Notably, T-TPU consistently exhibited higher temperatures than C-TPU across all tested bands, attributable to its superior optical density in the 300 nm to 1000 nm range. Interestingly, the equilibrium temperatures for both the UV and IR bands were higher than those for the visible band. Specifically, in the IR group (**Figure 3f**), the T-TPU layer reached significantly higher temperatures compared to the C-TPU layer, with temperature differences summarized in **Figure 3g**. The estimated temperature difference in the IR group was approximately 2.5 °C, that was comparable to the reported value of ~2.9 °C of ITO nanoink,[34] contrasting with ~0.5 °C differences observed in both the visible and UV groups. This highlighted the sensitivity of T-TPU to IR radiation, supported by a 3.59-fold greater absorbance in the IR region compared to C-TPU.

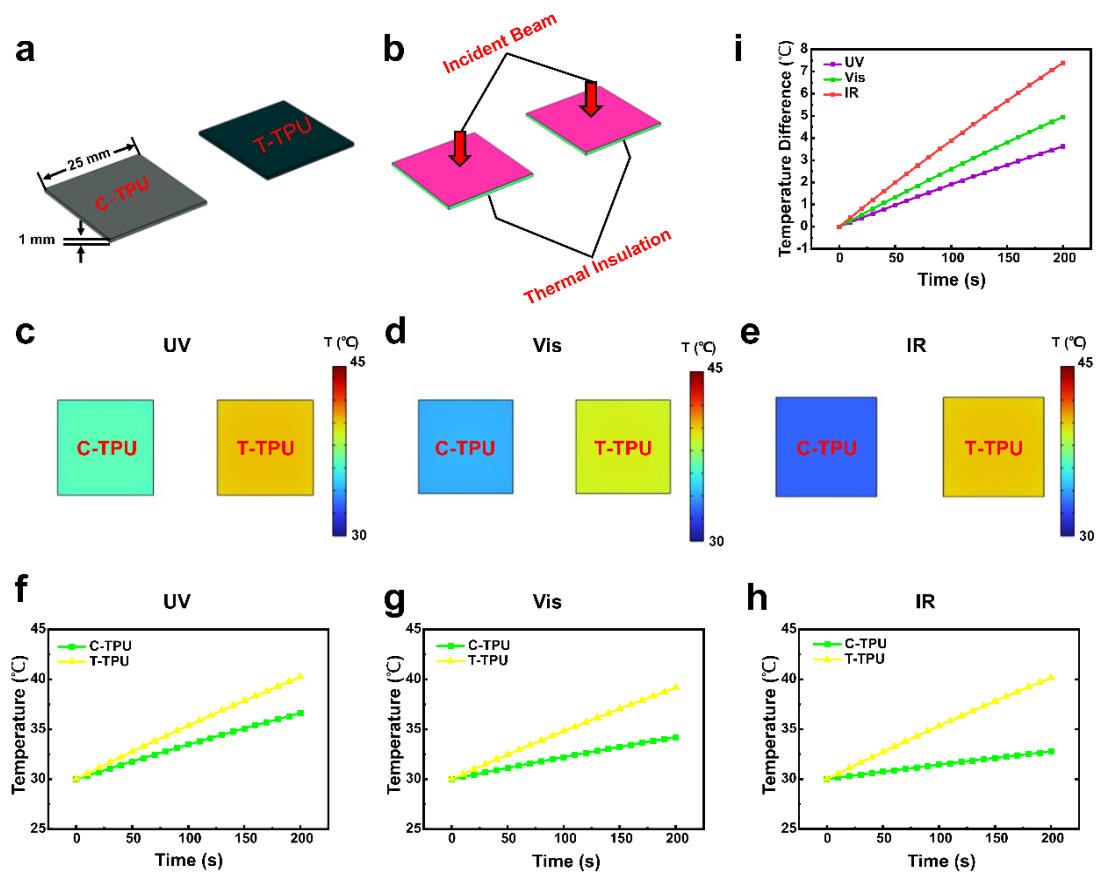

**Figure 4** Finite element method (FEM) simulation of the photothermal performance of C-TPU and T-TPU layers at various wavelengths. **(a)** Modeled C-TPU layer and T-TPU layer; **(b)** Boundary conditions of C-TPU and T-TPU layer; **(c) – (e)** Simulated temperature distributions of both layers under UV, visible (Vis), and infrared (IR) illumination over a duration of 200 seconds, respectively; **(f) – (h)** Summary of the temperature profiles of the C-TPU and T-TPU layers in response to UV, Vis, and IR illumination as functions of time; **(i)** Temperature differences between the C-TPU and T-TPU layers under UV, Vis, and IR illumination over the same time period.

To elucidate the structure-effect relationship between the optical density of the composites and the temperature elevation of the layers, a finite element method (FEM) numerical simulation was conducted. As illustrated in **Figure 4a**, both the C-TPU and T-TPU layers were modeled as square blocks with dimensions of 25 mm × 25 mm × 1 mm. The boundary conditions were detailed in **Figure 4b**, where the upper surfaces of the layers were subjected to incident beams with an input power of 0.15 W/cm², the lower surfaces were thermally insulated, and the upper surfaces along with the sidewalls were exposed to natural convective cooling heat flux from the external environment. The material parameters, listed in **Table S2**, were assigned to the respective domains, and the photothermal simulation was executed using the heat transfer module, which incorporated coupled interfaces for solid and radiative beams in absorbing media to compute the temperature distribution and changes over a duration of 0 to 200 seconds. **Figure 4c - 4e** presented the calculated temperature distributions for the C-TPU and T-TPU layers under 200 seconds of UV, visible (Vis), and infrared (IR) illumination, respectively. Notably, the T-TPU layer consistently exhibited higher temperatures than the C-TPU layer, with the IR illumination resulting in the highest temperature readings compared to UV and visible light, corroborating the trends observed in prior experiments. **Figure 4f - 4h** summarized the temperature variations for the C-TPU layer (green curves and dots) and the T-TPU layer (yellow curves and dots) under UV, visible, and IR light. Consistent with experimental measurements, the C-TPU layer achieved elevated temperatures of 36.6 °C, 34.2 °C, and 32.7 °C under UV, visible, and IR light, respectively, compared to the T-TPU layer, which recorded temperatures of 40.2 °C, 39.1 °C, and 40.2 °C. The corresponding temperature differences are illustrated in **Figure 4i**, revealing a temperature difference of 7.4 °C between the C-TPU and T-TPU layers in the IR group, surpassing the differences of 3.6 °C and 5.0 °C observed in the UV and visible groups, respectively. These findings suggested that the proposed simulation model effectively captures the structure-effect relationship between the optical density of the composites and the temperature elevation of the layers.

## 3. Mechanical Property

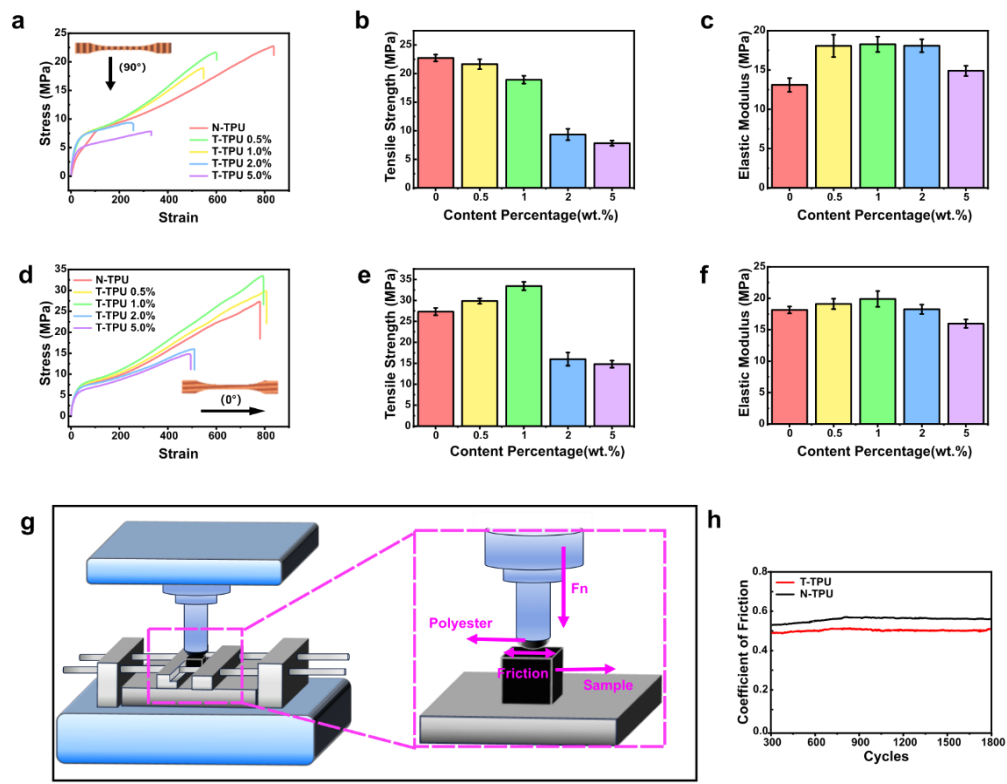

**Figure 5** Experimental characterization of the mechanical properties of T-TPU composite filaments. **(a) - (c)** Stress-strain curves, tensile strength and elastic modulus of T-TPU composite filaments series with printing stacking manners of [90°] direction; **(d) - (f)** Stress-strain curves, tensile strength and elastic modulus of T-TPU composite filaments series with printing stacking manners of [0°] direction; **(g)** Scheme diagram showed the coefficient of friction (COF) test setup; **(h)** Coefficients of friction of 5 wt% T-TPU and N-TPU samples measured over 1800 cycles.

The mechanical properties of composites were crucial, as they directly influenced the reliability, longevity, and range of applications of 3D-printed objects. Initially, tensile experiments were performed to assess the tensile strength and elastic modulus of T-TPU printed samples. As illustrated in **Figures S5a** and **S5b**, dumbbell-shaped samples were designed using CAD software and 3D printed with varying T-TPU filaments oriented in [0°] or [90°] for tensile testing. **Figure 5a** presented the stress-strain curves of T-TPU filaments printed with a [0°] stacking direction, while **Figure 5b** and **5c** summarized their tensile strength and elastic modulus, respectively. Notably, the tensile strength of the T-TPU composite decreased by over 50 % when the weight percentage of TiN exceeded 2 %. Conversely, the elastic modulus of T-TPU exhibited an enhancement greater than 30 % for TiN weight percentages below 2 %, but this improvement diminished as the weight increased to 5%. Similarly, **Figure 5d** displayed the stress-strain curves for samples printed with T-TPU filaments in a [90°] stacking direction, with the corresponding tensile strength and elastic modulus presented in **Figure 5e** and **5f**, respectively. For these samples, the tensile strength also demonstrated a significant decrease of over 40%, while the elastic modulus varied, peaking at 19.5 MPa with 1% TiN content. The uniform distribution of TiN nanoparticles within the TPU matrix effectively filled voids, increasing disorder within the matrix and restricting the movement of molecular chains, ultimately weakening the softness and ductility of the TPU. Additionally, wear resistance was evaluated using a reciprocating ball-and-disk friction wear tester, as depicted in **Figure 5g**. **Figure 5h** compared the coefficient of friction (COF) curves for the N-TPU and 5 wt% T-TPU surfaces, revealing that the T-TPU surface exhibited a lower and more stable COF of approximately 0.5, compared to about 0.53 for N-TPU. The uniform dispersion of TiN nanoparticles in the TPU matrix contributed to a lubricating effect that reduced the direct contact area between friction surfaces, thus lowering the COF. Remarkably, the low COF of the T-TPU surface withstood approximately 1,800 friction cycles, indicating excellent abrasion resistance.

In summary, the introduction of a small percentage of TiN nanoparticles led to improvements in mechanical properties, including tensile strength, elastic modulus, and abrasion resistance. However, the enhancements in tensile strength and elastic modulus diminished when the TiN content increased to 5 %.

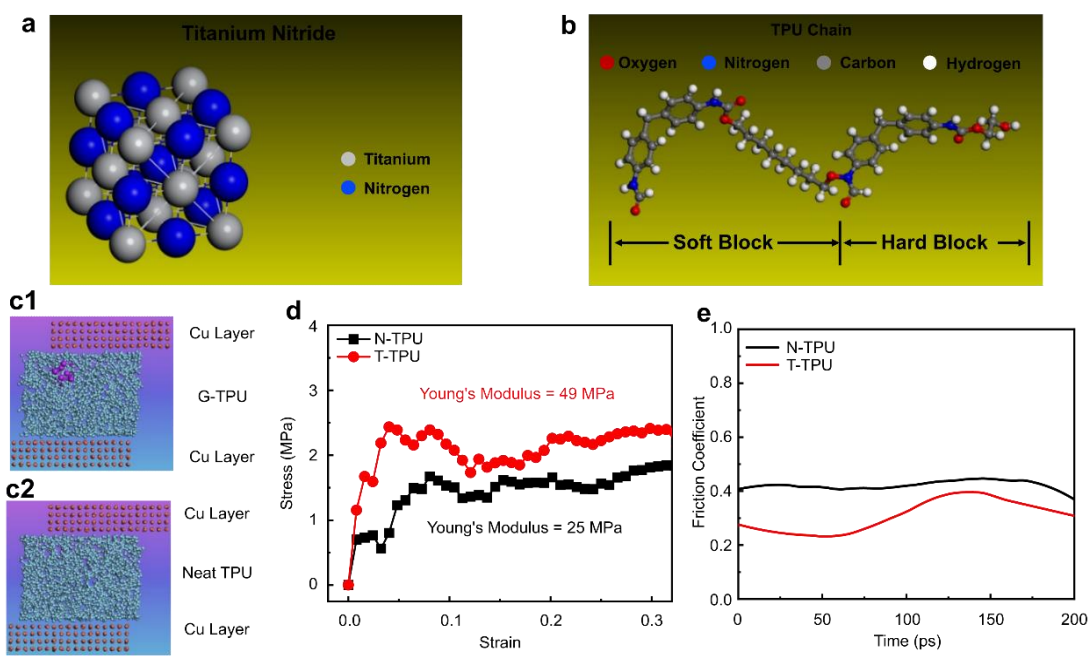

**Figure 6** Molecular dynamic (MD) simulation of the mechanical properties of T-TPU composite. **(a)** Modeled TiN lattice; **(b)** Modeled TPU repeating unit; **(c1)** Modeled Cu/T-TPU/Cu layer and **(c2)** Modeled Cu/N-TPU/Cu layer for confined shear simulation; **(d)** Calculated stress-strain curves and estimated Young's Modulus of N-TPU and T-TPU.; **(e)** Calculated COF of N-TPU and T-TPU surface.

To investigate the enhancement of TiN nanoparticles in the polymer composite at the atomic and molecular levels, molecular dynamics (MD) simulations were conducted using Materials Studio software. As illustrated in **Figure 6a**, the TiN lattice was characterized as a face-centered cubic (FCC) structure composed of alternating titanium and nitrogen atoms, followed by geometric optimization using the CASTEP module. The repeating unit of TPU, depicted in **Figure 6b**, was modeled with a molar ratio of soft block to hard block of 1:1 after geometric optimization with the Dmol3 module. Subsequently, boxes for T-TPU (5 wt% TiN) and N-TPU were constructed using the Amorphous Cell module, and then assembled into layers for mechanical property studies and confined shear simulations via the Forcite tools, as shown in **Figures 6c1** and **6c2**. The calculated stress-strain curves for N-TPU and T-TPU are presented in **Figure 6d**, demonstrating that T-TPU exhibited a higher Young's modulus (49.0 MPa) compared to N-TPU (25.0 MPa), which was in reasonable agreement with the experimental characterization shown in **Figures 5c** and **5f**. Additionally, **Figure 6e** evaluated the friction coefficients of N-TPU and T-TPU during the confined shear simulation, revealing that T-TPU (~0.32) exhibited lower friction interactions than N-TPU (~0.40), corroborating the experimental results presented in **Figure 5h**. These molecular dynamics simulation results confirmed that the incorporation of TiN nanoparticles enhances both the elastic modulus and lubrication properties, thereby supporting the potential for photothermal applications of the T-TPU composite.

## Conclusion

In this study, T-TPU composite filament was successfully synthesized, demonstrating excellent printability using a fused deposition modeling (FDM) 3D printer. Notably, T-TPU exhibited significant improvements in thermal properties, including enhanced thermal conductivity (measured in 0.52 W/(m·K)) for T-TPU with 5 wt% TiN compared to 0.26 W/(m·K) of N-TPU, increased heat capacity (1.99 J/(g·°C) of T-TPU with 5 wt% TiN versus 1.75 J/(g·°C) of N-TPU), and a slight increase in density (1.76 g/cm³ of T-TPU with 5 wt% TiN compared to 1.21 g/cm$^3$ of N-TPU), which was advantageous for its photothermal performance. As anticipated, solar photothermal de-icing and infrared labeling experiments indicated the exceptional photothermal capabilities of T-TPU. Specifically, T-TPU patterns exhibited a temperature approximately 2.5 °C higher than that of the commercial black TPU (C-TPU) patterns at equilibrium after 2 minutes of exposure to simulated sunlight at 0.15 W/cm², achieving over a 40 % reduction in photothermal de-icing time compared to C-TPU. Due to the plasmonic characteristics of TiN nanoparticles, T-TPU demonstrated a notable plasmonic response, with temperature differences of approximately 2.5 °C compared to C-TPU patterns under infrared (805 nm, 0.10 W/cm²) illumination, whereas the differences under UV (360 nm, 0.10 W/cm²) and visible (525 nm, 0.10 W/cm²) radiation were less pronounced at approximately 1.0 °C. Finally, the enhancement of mechanical properties in the T-TPU composite was evaluated. Overall, the introduction of a small percentage of TiN nanoparticles resulted in improvements in mechanical properties, including tensile strength, elastic modulus, and abrasion resistance, thereby supporting potential applications in demanding environments.

## Experimental Section

Essential Experimental Procedures/Data are available in the Supporting Information of this article.

## Supporting Information

The Supporting Information is available free of charge on the ACS Publications website at DOI:XXXXX

## Authorship contribution statement

**Miss Siyu Lu:** Data curation, Writing- Original draft preparation. **Dr. Min Xi:** Conceptualization, Methodology, Software, Writing- Reviewing and Editing. **Prof. Jixiang Zhang:** Supervision, Writing- Reviewing and Editing. **Prof. Nian Li:** Supervision, Writing- Reviewing and Editing. **Prof. Zhenyang Wang:** Supervision, Writing- Reviewing and Editing, Project administration, Funding acquisition.

## Conflict of competing interest

The authors declare that they have no known competing financial interests or personal relationships that could have appeared to influence the work reported in this paper.

## Acknowledgements

This work was financially supported by the National Key Research and Development Project (2022YFA1203600, 2020YFA0210703), and the National Natural Science Foundation of China (Nos. 12204488), Joint Training Base Construction Project for Graduate Students in Chongqing (JDLHPYJD2020031), Shandong Province Science and Technology Small and Medium-sized Enterprise Innovation Ability Promotion Project (2024TSGC0948).

# Supporting Information

# 3D Printable Plasmonic Titanium Nitride Nanoparticles Enhanced Thermoplastic Polyurethane Composite for Improved Photothermal De-Icing and Infrared Labeling

Siyu Lu[a], Jixiang Zhang[a], Min Xi*[b], Nian Li*[b], Zhenyang Wang*[b]

[a] School of Mechatronics and Vehicle Engineering, Chongqing Jiaotong University, Chongqing, 400074, P. R. China

[b] Institute of Solid State Physics and Key Laboratory of Photovoltaic and Energy Conservation Materials, Hefei Institutes of Physical Science, Chinese Academy of Sciences, Hefei, Anhui 230031, P. R. China

Corresponding authors:

Min Xi − orcid.org/0000-0003-4414-3110; Email: minxi@issp.ac.cn

Nian Li; Email: linian@issp.ac.cn

Zhenyang Wang − orcid.org/0000-0002-0194-3786; Email: zywang@iim.ac.cn

**Experimental Methods**

**Materials**

Titanium nitride (TiN, >97%), N,N-dimethylformamide (DMF, AR, 96%) were provided by Shanghai Aladdin Biochemical Technology Co.,Ltd. Thermoplastic polyurethane (TPU) with a hardness of 95A was purchased from Toprene Enterprise Co., Ltd. Colored TPU (black) was purchased from Raised 3D Technology Co., Ltd.

**Preparation of 3D Printing Filaments.**

The preparation of T-TPU filament involved a two-step mixing procedure, comprising solution mixing and melt blending, with specific modifications.[1] Initially, TiN nanoparticles and TPU granules were melt blended in varying ratios (0 wt%, 0.5 wt%, 1 wt%, 2 wt%, and 5 wt%) using a WLG 10 G twin-screw extruder (Shanghai Xinshuo Precision Machinery Co., Ltd., China). This process ensured that the TiN nanoparticles were uniformly dispersed within the TPU matrix. Following the melt blending, the resulting rough T-TPU composite was extruded and manually cut into masterbatch pellets after cooling. Subsequently, the masterbatch was processed into wire-shaped filaments utilizing a D12 single-screw extruder (Shanghai Xinshuo Precision Machinery Co., Ltd., China). The extrusion parameters were carefully regulated, with the screw speed set at 30 rpm and the temperature maintained at 200 °C. Ultimately, the composite filaments, featuring a diameter of 1.75 mm, were produced for use in fused deposition modeling (FDM) printing.

**Preparation of 3D Printed Samples.**

Prior to printing, the filaments were dried in an oven at 60 °C for 5 hours. The printing process was conducted using a Raise 3D Pro 3 3D printer, equipped with 0.4 mm diameter nozzles. The 3D models were designed using SolidWorks software and subsequently sliced with ideaMaker software. All printing parameters were established and controlled through ideaMaker, as detailed in **Table S1**. Notably, two stacking modes, [0°] and [90°], were employed, with the term "stacking mode" denoting the angle between the melt filling direction and the printing direction. Various print paths and stacking methods were configured in the ideaMaker software, maintaining a print temperature of approximately 230 °C, a print speed of 30 mm/s, and a platform temperature of 60 °C. The layer thickness was set to 0.1 mm, and the infill density was established at 100 % to ensure a dense and refined structural composition for all samples.

Table S1 The main FDM printing parameters

| Printing parameters | Value |
| --- | --- |
| Nozzle temperature (°C) | 230 |
| Platform temperature (°C) | 60 |
| Left nozzle diameter (mm) | 0.4 |
| Right nozzle diameter (mm) | 0.4 |
| Layer height (mm) | 0.1 |
| Filling rate (%) | 100 |
| Printing speed (mm/s) | 30 |
| Flow compensation | 1.2 |
| Stacking manner (°) | [0°] or [90°] |

**Morphological Characterization.**

For the characterization of TiN nanopowders, the sample was thoroughly dispersed in deionized (DI) water after 5 minutes of ultrasonication. Subsequently, it was transferred onto a copper sample grid for transmission electron microscopy (TEM) analysis using a JEM-2100F microscope (Japan) operated at an accelerating voltage of 160 kV, as illustrated in **Figure S1**.

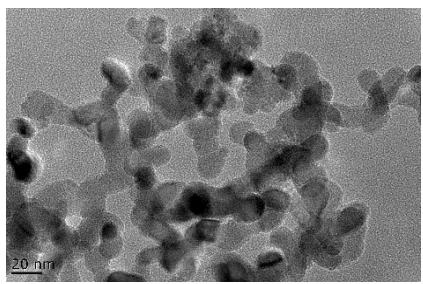

**Figure S1** TEM images of TiN nanopowder.

For the characterization of the T-TPU filament composition, the composite filament was cut into thin slices approximately 100 nm thick using an ultra-thin nano-blade slicer. The sliced samples were then carefully transferred onto a copper sample grid for analysis via transmission electron microscopy (TEM, JEM-2100F, Japan) at an accelerating voltage of 160 kV.

To examine the morphology of the T-TPU filament, the extruded filament was fractured after being cooled in liquid nitrogen and subsequently loaded onto a stage for scanning electron microscopy (SEM) analysis (SU8020, HITACHI, Japan) at an acceleration voltage of 10 kV.

**XRD Characterization.**

The crystalline structure of titanium nitride was analyzed using an X-ray diffractometer (SmartLab SE, Rigaku, Japan) equipped with Cu Kα radiation ($\lambda = 0.15406$ nm). This analysis confirmed the successful incorporation of titanium nitride nanoparticles into the TPU matrix and the purity of the filler. The scanning angle was set between 10° and 80°, with a scanning speed of 10°/min. The voltage and current were maintained at 40 kV and 100 mA, respectively.

**Thermogravimetric Analysis (TGA).**

Thermogravimetric analysis (TGA, TGA2, METTLER TOLEDO, Switzerland) was conducted to evaluate the composition and thermal stability of the various composites. Approximately 10 mg of each sample was heated in a nitrogen atmosphere at a rate of 10 °C/min, spanning a temperature range from room temperature to 800 °C. The residual weights of the different samples were subsequently recorded, as illustrated in **Figure 1d**.

**Thermal Conductivity Measurement.**

To assess the thermal conductivity of the T-TPU layer, a sample was fabricated in the form of a plate model, measuring 20 × 20 × 2 mm³, as depicted in **Figure S2**. The thermal conductivity was evaluated in the through-plane direction, indicated by the red arrow.

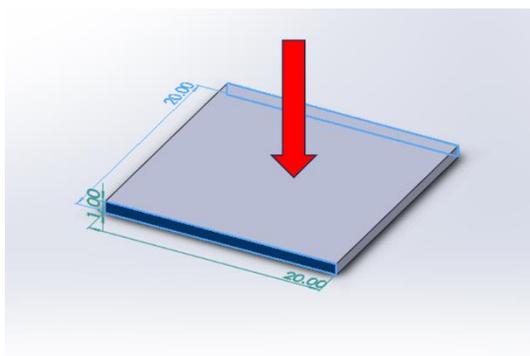

**Figure S2** Layer sample prepared for thermal conductivity measurements with model dimensions noted and red arrows indicating the direction of thermal conductivity.

The thermal conductivity was calculated as: TC = $\alpha C_p \rho$, where α is the thermal diffusivity, $C_p$ is the specific heat capacity and ρ is the density of the material under study.[2] Shown in **Figure S3a**, the thermal diffusivity was measured by laser flash method (LFA 457, NETZSCH, Germany). ρ was measured using the drainage method based on Archimedes' principle (**Figure S3b**). The specific heat capacity was obtained by a differential scanning calorimeter (DSC25, TA Instruments, America) with a ramp rate of 5 °C/min over a temperature range of -10 ~ 50 °C (**Figure S3c**). And the calculated thermal conductivity was shown in **Figure S3d**.

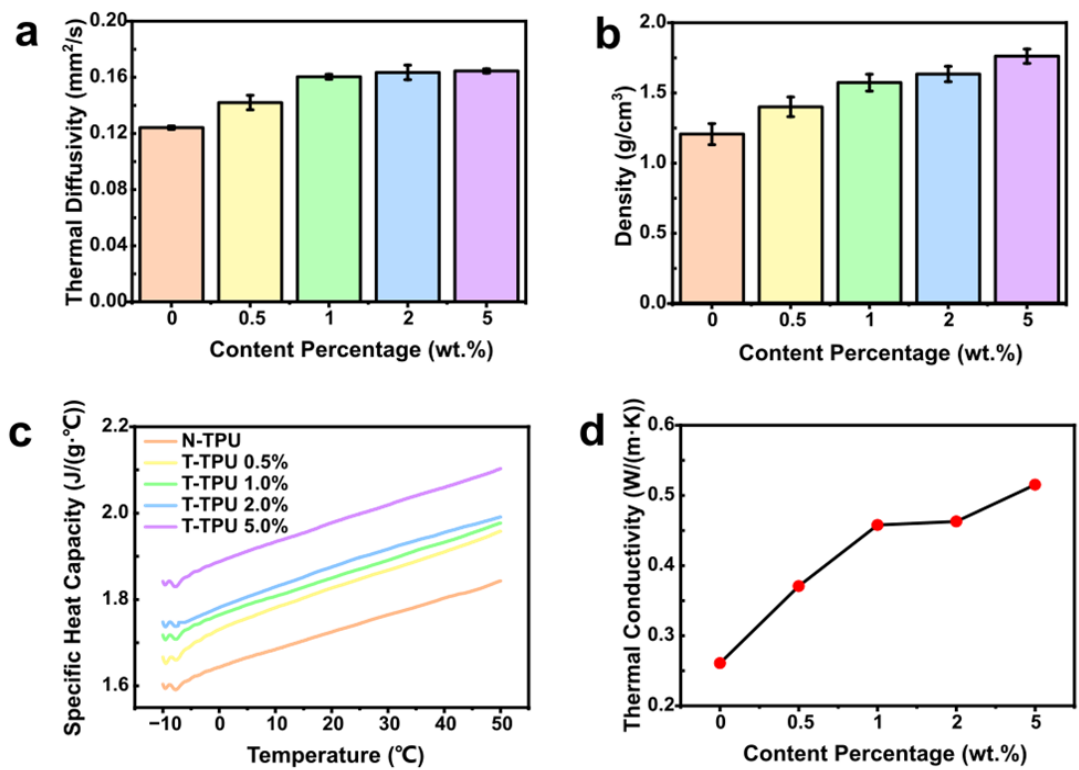

**Figure S3 (a)** Thermal diffusivity, **(b)** density, **(c)** specific heat capacity and **(d)** thermal conductivity of T-TPU filament with different weight percentage of TiN nanopowder.

**UV-Vis-NIR Spectrophotometer Measurement.**

To evaluate the transparency of the printed films, T-TPU, C-TPU, and N-TPU films with dimensions of 20 × 20 × 1 mm³ were prepared, as illustrated in **Figure 2c**. Absorption spectra in the wavelength range of 300 to 1000 nm were obtained and analyzed using a UV-Vis-NIR spectrophotometer (SOLID 3700, Shimadzu, Japan).

As depicted in **Figure 3c**, to quantify the optical density of the TPU filaments, T-TPU and C-TPU filaments were dissolved in DMF solvent at a concentration of 2 mg/mL for measurement. The resulting solutions were characterized using a UV-Vis-NIR spectrometer (UV-3600, Shimadzu, Japan).

**Photothermal Test**

In the photothermal infrared imaging test, the letter samples were printed according to the designed model in **Figure S4**. In specific, the letter samples of "C" and "T" were printed with the application of T-TPU filament, while the letter samples of "Q" and "J" were printed with the application of C-TPU filament.

In the photothermal infrared imaging test, letter samples were printed based on the designed model shown in **Figure S4**. Specifically, the letters "C" and "T" were printed using T-TPU filament, while the letters "Q" and "J" were printed with C-TPU filament.

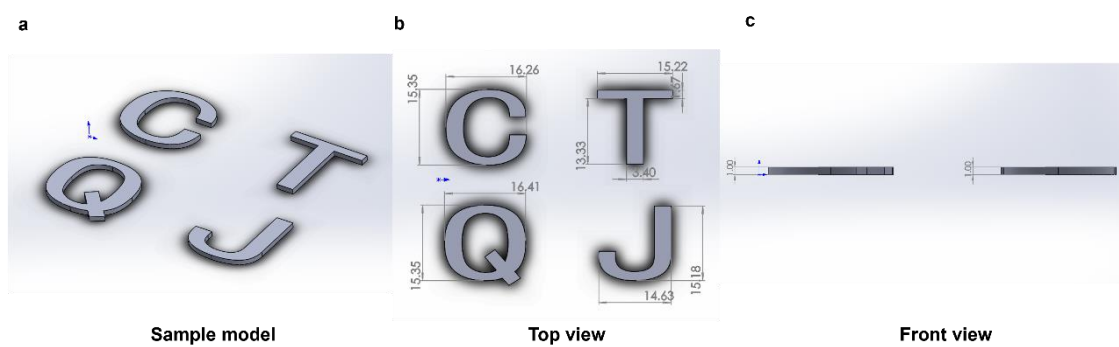

**Figure S4** Printed letter sample models **(a)** with model dimensions labeled in **(b)** top view and **(c)** front view.

Photothermal testing was conducted at an ambient temperature of approximately 22 °C. The printed letter samples were positioned on a sample stage and illuminated with a solar simulator set to an intensity of 0.15 W/cm². An infrared camera (FLUKE TiS55+) captured images of the samples every 10 seconds to monitor temperature distribution and variation.

**Finite Element Method (FEM) Simulations.**

The finite element method (FEM) simulation was conducted using the heat transfer module of COMSOL Multiphysics to investigate the photothermal performance of the 3D printed T-TPU and C-TPU layers under various LED light sources.

Two square plates, measuring 25 mm × 25 mm × 1 mm, were modeled based on the printed model files, as illustrated in **Figure 4a**. Material parameters (detailed in **Table S2**) were applied in the simulation. Notably, the absorption coefficients for the 5 wt% T-TPU and C-TPU were determined by quantifying the optical density of their dissolved solutions in the UV-Vis-NIR spectrum, as shown in **Figure 3c**. The boundary conditions were established as depicted in **Figure 4b**, with natural convective cooling heat flux applied to the upper layer and side walls, thermal insulation for the bottom layer, and illumination over the top region, delivering an input power of 0.15 W/cm² in the indicated pink area.

Table S2 The main parameters of the materials used for numerical simulation

| Materials | | 5 wt% T-TPU | C-TPU |
|---|---|---|---|
| Stacking Manners | | [0°, 90°] | [0°, 90°] |
| Thermal Conductivity | | Isotropic, 0.35 W/(m·K) | Isotropic, 0.24 W/(m·K) |
| Heat Capacity | | 1426 J/(kg·K) | 1500 J/(kg·K) |
| Density | | 1650 kg/m$^3$ | 1250 kg/m$^3$ |
| Absorption Coefficient | UV (360 nm) | 6.21 m$^{-1}$ | 4.07 m$^{-1}$ |
| | Vis (525 nm) | 5.55 m$^{-1}$ | 2.55 m$^{-1}$ |
| | IR (805 nm) | 6.15 m$^{-1}$ | 1.67 m$^{-1}$ |

The photothermal simulation was conducted using the heat transfer module, incorporating coupled interfaces for solid and radiative beams in absorbing media. The overall structure was meticulously meshed, resulting in a model comprising 1,000,000 degrees of freedom. This configuration enabled the calculation of temperature distribution and variations over a time span of 0 to 200 seconds.

**Mechanical Tests.**

Tensile tests were performed using a universal testing machine (AGS-X2kN, Shimadzu, Japan) in accordance with the ISO 37 standard. As illustrated in **Figure S5a**, the dumbbell-shaped samples had a thickness of 2 mm and a length of 75 mm. **Figure S5b** displayed the samples printed with varying stacking orientations, with a loading rate set at 50 mm/min. The elastic modulus was calculated within a strain range of 1 to 5 %. A total of five samples from each composite type were tested, and the average tensile strength and elastic modulus were subsequently computed.

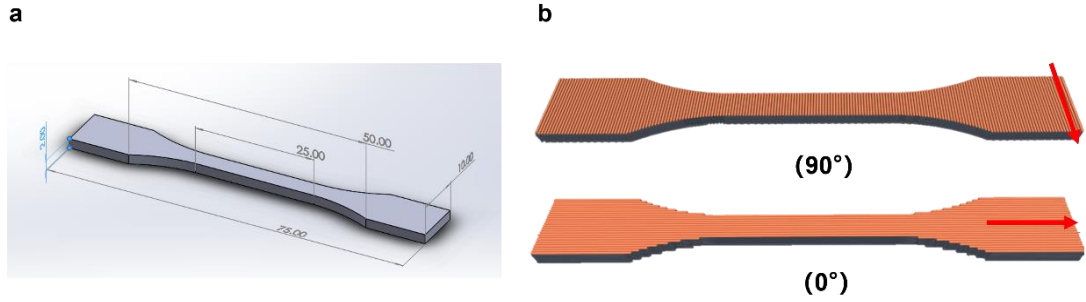

**Figure S5 (a)** Printed dumbbell-type samples for tensile testing sample models and model dimensions; **(b)** Schematic of the models deposited in two different stacking orientations, the upper image and lower image showed the sliced models with [90°] and [0°] deposition orientation of stacking, respectively.

The coefficient of friction (COF) was measured under ambient air conditions using an MFT-5000 friction and wear tester (Rtec Instruments, USA). Steel balls (Φ = 6 mm) served as the friction counter pairs, and a load of 2 N was applied. The test involved a reciprocating amplitude of 10 mm and a frequency of 1 Hz.

**Molecular Dynamics Simulations.**

The molecular dynamics simulations were performed by Materials Studio. The TiN lattice was modeled as face centered cubic of alternating titanium and nitrogen atoms (**Figure 6a**). The CASTEP module was applied to optimize the geometry of the lattice, and the resulting lattice parameters were calculated as: a = b = c = 4.25 Å, α = β = γ = 90°. And as shown in **Figure 6b,** the repeating unit of TPU was described as one soft chain block that developed from octanediol ($C_8H_{18}O_2$) and diethelene glycol ($C_2H_6O_2$) and one hard chain that developed from methylenediphenyl diisocyanate ($C_{15}H_{10}N_2O_2$) and diethelene glycol ($C_2H_6O_2$). The geometric optimization was performed with Dmol3 module for energy minimization.

The TPU matrix of filament was in molecular weight of ~1.5 million. To simplify the computation, the box of N-TPU or T-TPU were modeled by packing the TiN lattice and repeating units of TPU into a box to reach the density of ~1.3 g/cm³ or ~1.5 g/cm³ that closed to the experimental value. Specifically, the N-TPU box and T-TPU (5 wt% TiN) box with dimension of 28.9 Å × 28.9 Å × 23.0 Å were built by packing desired composition to reach defined density with Amorphous Cell module. The built box was then assembled into Cu/N-TPU/Cu layer or Cu/T-TPU/Cu layer, and 100,000 cycles annealing steps were performed for stabilization. Mechanical property studies and confined shear simulations were then carried out, utilizing the COMPASS II force field for accurate modeling.